\documentclass[preprint2]{aastex}
\usepackage{longtable}
\usepackage{natbib}

\shorttitle{UV Indices of SSPs}
\shortauthors{Chavez et al.}

\begin{document} \title{
Mid-UV Narrow-Band Indices of Evolved Simple Stellar Populations}

\author{M. Chavez$^{1}$, E. Bertone$^1$, J. Morales-Hernandez$^1$ and A. Bressan$^{1,2,3}$}
\affil{$^1$INAOE, Luis Enrique Erro 1, 72840, Tonantzintla, Puebla, Mexico}
\email{mchavez@inaoep.mx, ebertone@inaoep.mx, jmorales@inaoep.mx}
\affil{$^2$INAF, Osservatorio Astronomico di Padova, Vicolo dell'Osservatorio 5, Padova, Italy}
\affil{$^3$International School for Advanced Studies-SISSA, Strada Costiera n. 4, Trieste, Italy}
\email{alessandro.bressan@oapd.inaf.it}

\begin{abstract}
We explore the properties of selected mid-ultraviolet (1900--3200~\AA)
spectroscopic indices of simple stellar populations (SSPs). We incorporate the
high resolution UVBLUE stellar spectral library into an evolutionary
population synthesis code, based on the most recent Padova isochrones.
We analyze the trends of UV indices with respect to age and chemical
composition. As a first test against observations, we compare our results
with the empirical mid-UV spectral indices of Galactic globular clusters, 
observed with the {\it International Ultraviolet Explorer} (IUE).

We find that synthetic indices exhibit a variety of properties, the main one
being the slight age sensitivity of most of them for ages $>$ 2~Gyr. However,
for high metallicity, two indices, Fe~\textsc{ii}~2332 and Fe~\textsc{ii}~2402, display a remarkably
different pattern, with a sharp increase within the first two Gyr and,
thereafter, a rapid decline. These indices clearly mark the presence of young
($\sim$1~Gyr) metal rich ($Z \geq Z_\odot$) stellar populations.

We complement existing UV indices of Galactic globular clusters with new
measurements, and carefully identify a sub-sample of ten indices suitable for
comparison with theoretical models. The comparison shows a fair agreement and,
in particular, the strong trend of the indices with  metallicity is well
reproduced.

We also discuss the  main improvements that should be considered in future
modelling concerning, among others, the effects of $\alpha$-enhancement in the
spectral energy distributions.
\end{abstract}


\section{Introduction}

This is the third paper of a series devoted to the analysis of the ultraviolet
(UV) morphology of intermediate and late type stars and evolved stellar
populations. In the first paper \citep[hereafter Paper~I]{lino05}, we
presented the UVBLUE library of synthetic stellar
spectra at high resolution and provided its potential usefulness and
limitations, when applied to complement empirical data in the analysis of
stellar spectra. In a second paper \citep[hereafter Paper~II]{chavez07}, we
applied UVBLUE to the detailed analysis of a set of 17 mid-UV spectroscopic
indices in terms of the main atmospheric parameters: effective temperature,
surface gravity, and metallicity. This study also included the comparison of
synthetic indices with those measured in {\it IUE} spectra of a sample of main
sequence stars. The global result was that, with UVBLUE properly degraded to
match the resolution of {\it IUE} (6~\AA), eleven synthetic indices
(absorption or continuum) either well reproduced the observations or could be
easily transformed to the {\it IUE} system. In this work, we further extend the
investigation to the simplest aged stellar aggregates: the globular clusters.

Globular clusters represent the classical test bench for any population
synthesis approach of the analysis of larger stellar aggregates in the local
as well as the distant universe \citep{schiavon07}. They are commonly
considered the prototypes of a coeval, chemically homogeneous stellar
population, and are used to build up the properties of elliptical galaxies.
Although the integrated ultraviolet light of globulars has been extensively
analyzed, the study of their UV morphology has not yet been fully exploited
spectroscopically through the analysis of their absorption indices
\citep{fanelli90,fanelli92}.  This analysis provides
numerous advantages over the full UV spectral energy distribution (SED), since
indices are easily calculated and generally defined in narrow bands,
diminishing substantially the effects of interstellar reddening and flux
calibration. Chemical species can be examined separatedly, at
least on those indices dominated by a single atom (i.e., no
blends).

Over the past decade Fanelli's et al.\ absorption indices have been
incorporated in a variety of analyses. \citet{ponder98}, for instance,
analyzed a sample of four globular clusters in M31 and six elliptical galaxies
observed with the Faint Object Spectrograph on board the {\it Hubble Space
  Telescope}. They compare these data, in the form of colors and line indices,
with a sample of Galactic stars, Milky Way globular clusters, and the
integrated UV spectrum of M32. They found that Milky Way systems, clusters in
M31, and elliptical galaxies conform three distinct stellar populations. Almost
contemporarily, \citet{rd99} carried out a detailed comparative analysis of
M32 and NGC~104 (47~Tuc) aimed at constraining the stellar content of M32. In view of
the chemical similarities (both nearly solar) they modelled the UV spectral
indices of M32 on the basis of the better studied Galactic stellar system, in
terms of its leading parameters (age and metallicity). In spite of their
compatible chemical composition, they found that, while spectroscopic indices of
NGC~104 can be interpreted in terms of a dominant main sequence (turn-off) and
various contributions of other stellar types, in particular red objects in the
horizontal branch (HB), M32 requires a significantly more
metal rich turn-off (TO) and a more important contribution
of stars of type A, likely consistent with the observed
paucity of post-asymptotic giant branch stars \citep{bertola95,brown08}.
More recently, \citet{lotz00}
conducted an analysis of the effects of age and chemical composition on the
integrated spectra of simple populations modelled with population synthesis
techniques. They tested and used as ingredients the library of theoretical
fluxes by \citet{Kurucz93} and a modified set of the isochrones of
\citet{bertelli94}; they found that, within the limitations of Kurucz's low
resolution SEDs (see Paper II), which appeared to faithfully reproduce the
stellar indices BL~2538 and S2850, simple population models adequately match
the mid-UV morphology of mean globular groups \citep{bba95}. However, they
failed to match the mid-UV indices of galaxies, ascribing the discrepancies,
among other possible agents, to the presence of a hot stellar component [e.g.,
blue straggler stars (BS)] and a mixture of metal-poor objects (and a
metal-rich young population, as in the case of M32).

It is fair to mention that the above cited papers provide some of the main
drivers of the analysis presented in this paper: a)- Galactic globular
clusters (GGCs) has represented, with a few exceptions, the best reliable set of
single generation populations, therefore, any stellar population modelling
should be tested against the observed integrated properties of globulars; b)-
current empirical stellar libraries still present a marked paucity in the
parameter space, particularly in chemical composition. Such a deficiency has
been managed with the use of either grids of theoretical spectra
\citep{lotz00} or by extending the stellar database with a set of stellar
spectra of metal-rich stars with fiducial chemical compositions
\citep{rd99}. Nonetheless, it was until recently that theoretical databases at
the appropiate resolution to handle, for instance, {\it IUE} spectroscopic data became
available. Additionally, it has been suggested that the mid-UV properties of
Milky Way systems might not be applicable to the study of extragalactic
evolved populations \citep{ponder98}. It is therefore imperative that we test
theoretical predictions with local 
globular clusters and establish the necessary corrections, before we can rely on synthetic populations 
for studying non-local or more complex stellar systems.

Preliminary results on the application of synthesis techniques to the study
of the UV morphology of old populations have been already presented in
\citet{lino04}, \citet{bertone07}, and \citet{chavez08}. More recently
\citet{maraston08} have made use of our previous results to investigate
the far-UV and mid-UV indices in young populations with  the goal of providing
the tools for the study of distant post-starburst galaxies
(age $<$ 1~Gyr). Being based on the same theoretical and empirical framework, it
turns mandatory to at least briefly explain the ``added value" of the present
analysis. First, this paper extends their analyses to older populations,
including the age interval (1--3~Gyr), which incorporates the so far determined
(but apparently still rather inconclusive) ages of red galaxies at redshifts
around $z=1.5$ \citep[see, for instance,][]{Spinrad97}. Second, we provide a
comparison with Galactic globular clusters as a tool for quantifying the
required corrective factors for the theoretical indices. Third, eventhough we
concentrate in ``canonical" evolution, we provide, for the first time, hints
of the effects of non-solar partitions on the integrated ultraviolet energy
distributions of evolved populations. Fourth, we include more indices, some of
which appear to more clearly segregate the effects of age and metallicity in
old populations.

Aimed at complementing the necessary tools for the study of old populations at UV wavelengths, we present in this paper the
calculation of the integrated SEDs of SSPs in the UV spectral interval. We first theoretically
explore, in \S~\ref{sec:syn_pop}, the properties of integrated spectral features (in the form of spectroscopic and continuum indices) in terms of the main population parameters: metallicity and age. We then look, in \S~\ref{sec:obs}, into the empirical correlations between indices and chemical composition in a sample of Galactic globular clusters observed by {\it IUE}. In this section we also show the comparison between synthetic and empirical indices. A brief discussion on the agents that could explain
discrepancies (and their corrections) is provided in \S~\ref{problems}. In \S~\ref{alpha} we demonstrate the importance of adopting the
suitable element partition in particular in low resolution theoretical fluxes. A summary and conclusions are given in \S~\ref{summary}.

\section{Synthetic Single Stellar Populations at UV Wavelengths}
\label{sec:syn_pop}

We have computed new integrated spectra of SSPs following the prescriptions
outlined in \citet{bcf94} and the later revision by \citet{bgs98}.
The main differences with respect to \citet{bgs98} SSPs are summarized below.

In the present paper we incorporate the new Padova
isochrones \citep{bertelli08}.
These isochrones allow the calculation of SSPs with
arbitrary chemical compositions, metallicities that range
from $Z=$0.0001 to $Z=$0.07
in the assumption of a solar partition for the heavy elements,
and He abundance within reasonable ranges for each metallicity.
The asymptotic giant branch (AGB) phase is treated according to \citet{marigo08}.
The new isochrones do not include the post AGB phase and,
since this phase is relevant for the UV properties of SSPs,
we have added it following \citet{bertelli94}.

In our calculations we have considered ages from 100~Myr to 16~Gyr and the
following chemical compositions: Z=0.05, Y=0.28; Z=0.02, Y=0.28; Z=0.008,
Y=0.25; Z=0.004, Y=0.25; Z=0.0004, Y=0.25; Z=0.0001, Y=0.25. The red giant
branch (RGB) mass-loss parameter is very relevant for the present paper,
because it fixes the stellar masses on the horizontal branch (HB) and, in
turn, its hot blue tail. It has been set to $\eta_{RGB}$=0.50
independently from the metallicity as suggested by
\citet{carraro96} and, more recently, by \citet{vanloon08}.
The same RGB mass-loss parameter
has been used to reproduce the mid infrared colours
of 47~Tuc and  of selected old metal rich ellipticals in
the Coma cluster \citep{clemens09}.

The youngest populations will not be of any use
in the analysis of globular clusters presented in what follows.
Nevertheless, their calculation will
allow to probe the behavior of metallic features since their very onset, and will certainly
be an important ingredient in the planned analysis of composite systems, such as M32, in
which a young stellar generation (including A-type stars) might be present.

In computing the integrated SEDs\footnote{The full sample of SEDs are
  available upon request from the author.} we have made use of the UVBLUE
high-resolution synthetic stellar
library\footnote{http://www.inaoep.mx/$\sim$modelos/uvblue.html} (Paper~I). We
have linearly interpolated within the spectral library in order to match the
parameters considered for the mass bins along the isochrones.

\begin{figure*}[!t]
\begin{center}
\resizebox{!}{!}{\includegraphics{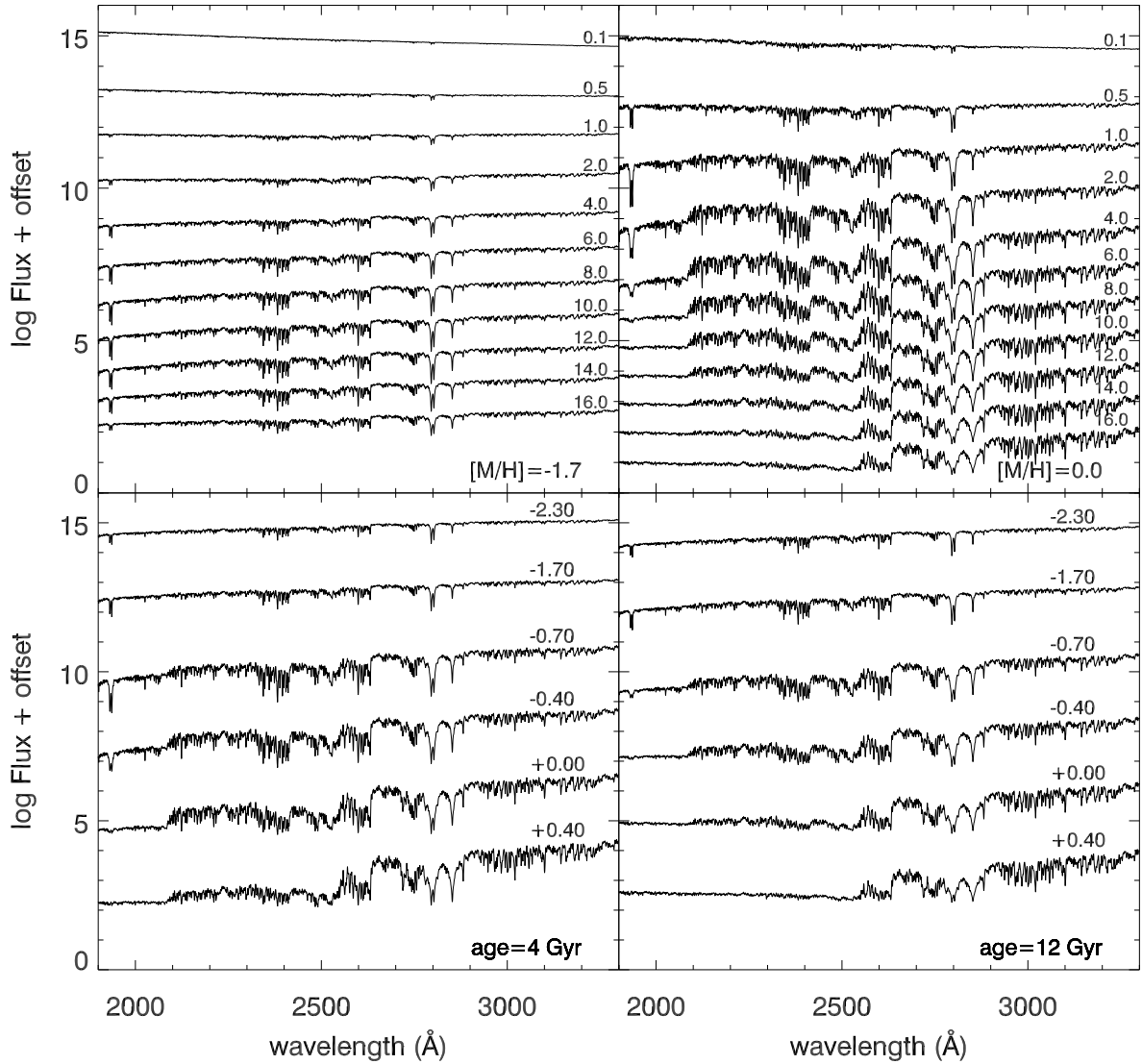}}
\caption{A sequence of synthetic spectral energy distributions for a set of metal-poor and solar metallicity populations is presented in the upper panels. The composite spectra include populations ranging from 100~Myr to 16~Gyr. The labels on the right indicate the age (in Gyr) of the
  population. Aimed at illustrating the effects of chemical composition we plot in the lower panels the sequence of integrated
fluxes for two representative ages, 4 and 12~Gyr, and six [Fe/H] values varying from $-2.30$ to $+0.40$.
\label{fig:seds_ages}}
\end{center}
\end{figure*}

Four sequences of integrated SEDs of SSPs are shown in panels of Fig.~\ref{fig:seds_ages}. In the upper panels, the sequences illustrate the effects of age in two synthetic populations of [M/H]=$-1.7$ and 0.0~dex, whose ages range from 0.1 to 16~Gyr as indicated by the labels on the right.
In the lower panels, we portray the effects of chemical composition for two representative old populations of 4 and 12~Gyr. A simple examination of the figure allows us to easily identify
changes in many spectral features including faint ones, variations that would
have been hidden if we had used a low resolution spectral grid (at, say,
10~\AA\ resolution, as the commonly used Kurucz low resolution database of stellar fluxes). For instance,
most of the spectral features (including the Mg~\textsc{ii} doublet at 2800~\AA) at the metal-poor regime display a slight increase over
the age interval 2--14~Gyr, with an apparent turnover at the oldest populations. In contrast, at solar metallicity we clearly
see that some features sharply increase up to relatively young ages and then
show a shallow decline with age, disappearing at about 12~Gyr. An example of
this behaviour is the series of features in the interval 2400--2500~\AA\ (mainly due to Fe~\textsc{i},
Fe~\textsc{ii}, Ni~\textsc{i}, Co~\textsc{i}, and Mg~\textsc{i}) which are evident all the way to 16~Gyr at low metallicities, but absent in the more metal-rich counterparts.
The quantitative measurement of these variations in terms of the leading
population parameters (age and chemical composition)
is given in the following section.

\subsection{The Absorption Spectroscopic Indices} \label{sec:indices_ssp}

With the set of synthetic SEDs, we have calculated 17 spectroscopic indices,
whose definition is given in Paper II; most of them were introduced by
\citet{fanelli90}. The indices measure the absorption of the most prominent
features in the mid-UV spectra of intermediate and late-type stars. Briefly,
an index is defined through three wavelength bands, with the two side bands
defining a pseudo-continuum, which is compared to the flux in the central band
pass, very much like the popular Lick indices defined in the optical
\citep[e.g.,][]{Worthey94}. For the calculation, we have
broadened the synthetic spectra with a Gaussian kernel of FWHM=6~\AA\ to
simulate the resolution of {\it IUE} in low resolution mode, which has been
(an still is) the work horse for comparison with empirical data. This step is
important, since we have demostrated that resolution might have non-negligible
effects on indices defined with the narrower bands (Paper~II).

Another important aspect is that we have, in the purely theoretical approach
presented in this section, included also some of the bluer indices. Unlike
previous investigations, we opted to include them foreseeing a potential use
in the analysis of higher quality data collected by either large optical
telescopes (for high redshift objects) or planned spaceborn instruments
\citep{ana08}.

On what follows we present the effects of age and metallicity on each of the 17 indices. Partial results have been already presented in
\citet{lino04} and \citet{bertone07}.

\subsubsection{The Effects of Metallicity and Age}
\label{glob_age_met}

Amongst the most important properties of theoretical stellar libraries, such
as UVBLUE, is that synthetic stellar SEDs can be computed for just about any
combination of stellar parameters, allowing, in turn, the calculation of SSPs
for a wide variety of combinations of age and chemical
composition. Empirically, previous analyses have coped with incompleteness in
the metallicity space, particularly at the metal-rich end, by complementing
available data sets with archival UV data and a homogeneous set of chemical
compositions \citep{rd99}. Another possibility, which however will also
require a metal-rich stellar library, is to assemble an empirical sequence of
more metallic populations. Perhaps, a way to do this task is to extend the
chemical composition with the integrated spectra of old open clusters, but,
hosting a significantly smaller number of stars, the results would be subject
to stocastic processes.

\begin{figure*}[!t]
\begin{center}
\includegraphics[height=16.0cm,angle=90]{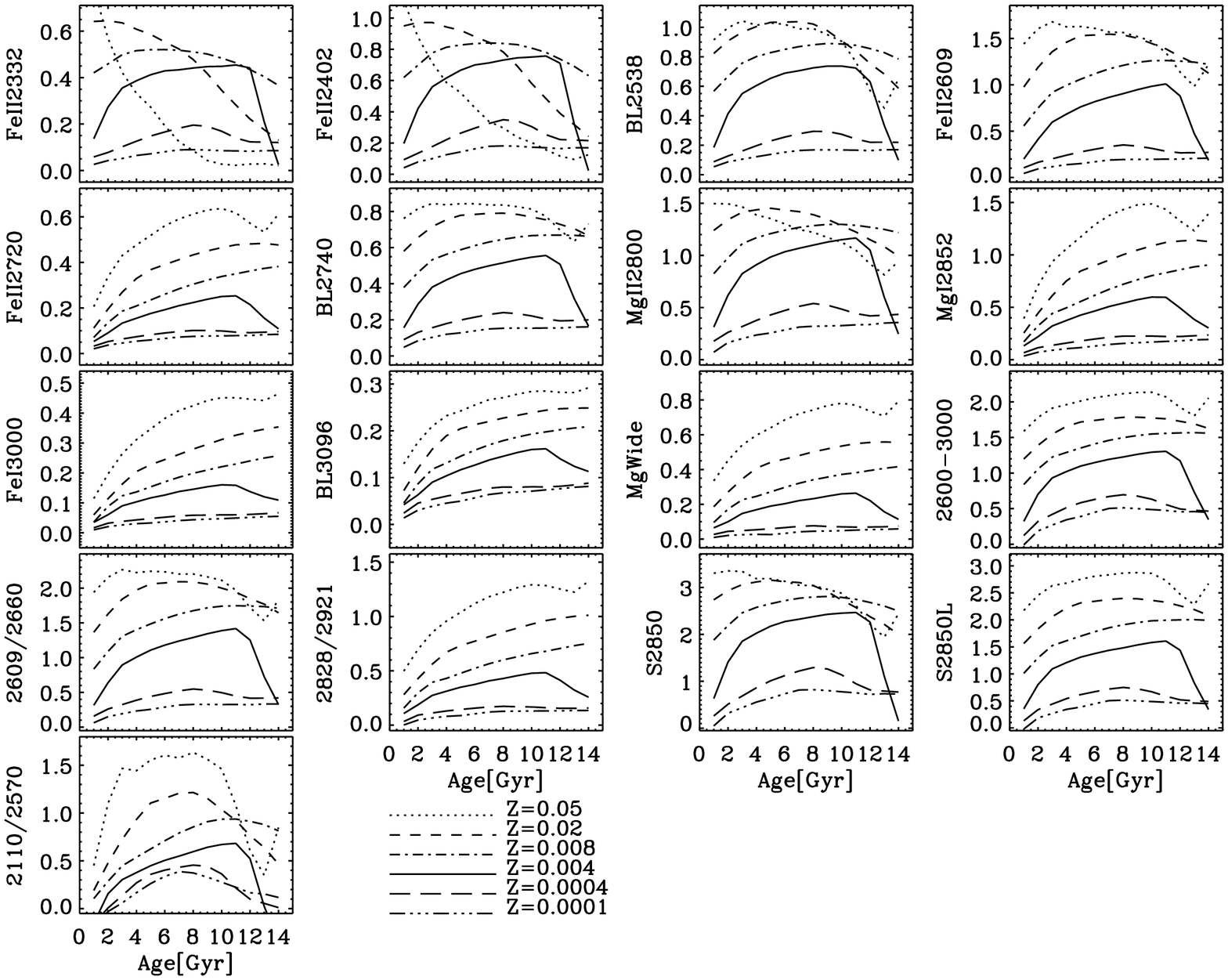}
\caption{Full set of mid-UV SSP synthetic indices as a function of age.
The different line types stand for different chemical compositions and helium contents:
Z=0.05,Y=0.28; Z=0.02,Y=0.28; Z=0.008,Y=0.25;
Z=0.004,Y=0.25; Z=0.0004,Y=0.25;
Z=0.0001,Y=0.25. Indices are in magnitudes.
\label{fig:seds_indexteffage}}
\end{center}
\end{figure*}

The combined effects of age and chemical composition are displayed in
Fig.~\ref{fig:seds_indexteffage}.
Taking into account that at ages less than 1~Gyr all the
absorption features are diluted by the strong continuum,
we may summarize the following behavior:

i) There are indices that sharply increase to their maximum
value within the first two Gyr and, thereafter,
remain almost constant up to 10~Gyr.

ii) At ages above about 11~Gyr some of the indices in the
most metal-rich populations show a more or less pronounced decline.
This is particularly evident in the case of $Z=0.004$
and likely reflects a combination of age, metallicity,
helium content, and mass-loss parameter that favour the
rapid development of a hot horizontal branch.
In some cases (e.g., Mg~\textsc{ii}~2800, S2850) the index value is
comparable to that of very low metallicity populations.

iii) At solar and super-solar metallicities, the two bluer indices,
Fe~\textsc{ii}~2332 and Fe~\textsc{ii}~2402, behave quite differently with
respect to the other indices: after a rapid rise (1~Gyr), there is a
relatively fast decline, which is steeper for the highest $Z$. 
The distinctive behavior of these two indices can be understood by examining
their behavior in stellar data displayed in Fig.~3 of Paper~II (see also
Fig.~2 of that paper). The stellar patterns are directly reflected in the
composite spectra of metal-rich SSPs, whose mid-UV emission is dominated by
turn-off stars. In stars, Fe~\textsc{ii}~2332 and Fe~\textsc{ii}~2402 are the
mid-UV indices that reach the maximum at the highest effective
temperature. Since they peak at about 7000~K at super-solar metallicity, in a
stellar populations the maxima of these 
indices are expected to take place in systems where turn-off stars are less
than 2~Gyr old. A similar reasoning is applicable to the solar chemical
composition case. The index peak $T_{\rm eff}$ is decreasing with
decreasing metal abundance, therefore in the less metallic populations the
maxima are reached at older ages. 
We have furthermore analyzed the effect of the dominant species in the band
passes of the three indices  Fe~\textsc{ii}~2332, Fe~\textsc{ii}~2402, and
Fe~\textsc{ii}~2609: we found that, whilst in fact ionized Fe has a dominant
role in their central band passes, at the same time it also significantly depresses
the blue band of Fe~\textsc{ii}~2609, making this index to be
less sensitive than the bluest two to Fe~\textsc{ii} abundance and to
effective temperature.

These indices mark the presence of a young metal-rich population, teherfore,
might be relevant for the study of high-redshift galaxies and/or to trace
rejuvenation episodes in the local old systems.

iv) Finally, another remarkable feature is that, with the
exception of the indices mentioned in (iii), the metallicity, not the
age, appears to be the main agent regulating the strength of most of
the indices, at least in the age interval 2--10~Gyr.


\section{UV Indices of Galactic Globular Clusters}
\label{sec:obs}

In this section, we present the
comparison of synthetic indices with those measured in a sample of globular
clusters observed by the {\it IUE}. The goal of the comparison is two-fold. On
the one hand, we want to explore the sensitivity of the indices to metallicity
from an empirical point of view, complementing the results of \citet{rd99}. On
the other hand, we want to test and discuss the capabilities of synthetic SSPs
to reproduce what is actually measured in observational data and, therefore,
identify fiducial features that can be modelled for the analysis of more
complex systems, which, due to the limitations of observed data sets, cannot
be studied from a purely empirical perspective.

We have extracted from the {\it IUE}-Newly Extracted Spectra (INES) database
the images of 27 globular clusters observed with the longwave prime and
redundant  cameras, large aperture and in low dispersion mode. The sample
represents the full set of GGCs observed with {\it IUE},
available in the INES data base. The working sample does not consider the
cluster NGC~5139 (Omega Cen) and  includes four more clusters than in the
work of \citet{rd99}. The analysis of NGC~5139\footnote{There are other GGC in
  our sample for which a composite main sequence has been identified
  (e.g. NGC~2808), nevertheless the abundance spread of iron peak elements 
turns out to be less pronounced that in Omega Cen.} will be
deferred to an on-going mid-UV study of the sample of early-type galaxies
observed by {\it IUE} being developed by our group. The unusual (as compared
to other globulars) composite nature in age and metallicity \citep[see,
e.g.,][]{sollima05,sdn06} of NGC~5139 will in fact serve as template for
testing combinations of SSPs. The four additional
clusters have spectra with a much lower quality than the rest, however, they
are still useful in some spectral intervals.

The sample is described in Table~\ref{glob_data}, where we list the cluster
identification, the iron abundance [Fe/H] from \citet{harris96}, the age
mainly from \citet{sw02}, the horizontal branch ratio\footnote{This quantity
  is defined as HBR = (B-R)/(B+V+R), where B, V, and R are the numbers of blue
  HB, RR-Lyrae, and red HB stars \citep{lee90}.} compiled
by \citet{harris96},
the horizontal branch type according to \citet{dickens72}, and, in the last
column, the additional references for the age not included in
\citet{sw02}. For two clusters, NGC~6388 and NGC~6441, which have been of
particular interest, because they represent a typical second parameter case,
we were not able to find in the literature age determinations. Based on the
evident similarities between NGC~6388 and NGC~104, \citet{catelan06} point out
that these clusters should be of similar age. Similarly, \citet{cd07} assumed
a reference age of 11~Gyr for NGC~6441.

It is worth to mention that sixteen objects have {\it IUE} images for which
the quality code (ECC) is four, five or six in the first digit. This, in
principle, would indicate {\it bonafide} data. This quality designation does
not guarantee, however, that the full wavelength range is useful. We have to
keep in mind that most globulars are instrinsically faint at the shortest
wavelengths and that the  sensitivity curves for both the LWP and LWR cameras
peak at about 2750~\AA\ and significantly degrade in the bluest region as well
as in the red, although in this latter region (say at 3000~\AA) the instrinsic
flux is much larger. These facts have imposed some limitations, which have
prevented in most analyses the use of the complete set of indices. For these
reasons, we have restricted our study to a subsample of ten
indices. The criteria for such a selection is somewhat complicated due to the
mixed instrumental effects and stellar theory limitations. However, within the
goals of this pioneering comparison between synthetic and empirical
integrated indices, we decided to start with indices that accomplish the
following criterion: we include narrow line indices, whose full bands are
defined in the interval 2400--2950~\AA, and broader line and continuum indices
in the range 2400--3130~\AA, hence excluding the indices Fe~\textsc{ii}~2402,
Fe~\textsc{i}~3000, BL~3096, and 2110/2570. Additionally, we have also excluded the
two indices BL~2720 and BL~2740 for which in Paper~II we found, respectively, that indices are largely overestimated by UVBLUE-based measurements and that the saturation of empirical indices is not reproduced by UVBLUE. Regarding this latter point, it should be noted that there are other indices that also get saturated (those displaying a hook-like behavior in Fig.~6 of Paper~II), nevertheless they differ from BL~2740 in an important aspect: the deviation from the linear correlation in this index takes place at values that would correspond to much younger ages than those expected for GGCs; therefore, age effects will blur
any existing correlation with chemical composition and any potential
compatibility with synthetic SSPs based on UVBLUE.

Note that we have also left the index Fe~\textsc{ii}~2332 in spite of not fulfilling the selection criteria.
Though unavoidably affected by the {\it IUE} sensitivity constraints,
we considered this feature important essentially because the synthetic analysis has shown that
the index displays remarkable different behavior with respect to the other
indices, turning it into potential tool for breaking the age-metallicity
degeneracy \citep{buzzoni08}. This index was measured in a restricted sub-sample of objects 
(15 out of 27), whose images are labelled with high quality codes.

In Table~\ref{glob_index}, we report the ten selected indices that have been
measured after correcting globular spectra for extinction, using the \citet{ccm89}
extinction curve, and for radial velocity. The error figures for five indices; namely 2600-3000, 2609/2660, 2828/2923, Mg~\textsc{ii}~2800, and Mg~\textsc{i}~2852 have been given by \citet{rd99}, based upon several images of individual clusters. We have independently estimated through an iterative process the typical
error associated with each index. The process consists of randomly adding artificial noise to each original spectrum and computing the index for each iteration. The standard deviation of the resulting (Gaussian) distribution of indices was taken as the error. Our estimates are in agreement with the typical values of \citet{rd99} of 0.03 (and 0.06 for Mg~\textsc{ii}~2800), except for Mg~Wide and Fe~\textsc{ii}~2332 for which we derived 0.01 and 0.05 mag, respectively. Empty entries in Table~\ref{glob_index} are
index values (mainly negative ones) that have been dropped after a detailed
visual evaluation of the quality of spectra in the bands defining each index.

We would like to point out an additional comment on the interstellar
extiction. Eventhough we have mentioned that spectroscopic indices do not
significantly depend on reddening, some of the indices cover up to 660~\AA\
and, bearing in mind that interstellar extinction dramatically increases at
mid-UV and shorter wavelengths, we consider it necessary to correct {\it IUE}
spectra for the effects of extinction. In fact, we conducted a test in which
we measured the set of indices, whose bands are defined at $\lambda >
2300$~\AA\ in the coadded spectrum (extracted from four {\it IUE} images) of
the slightly reddened cluster NGC~104. After correcting the spectrum, assuming
color excesses in the interval 0.0 to 0.8~mag (at a step of 0.1~mag), which
roughly correspond to the values comprised in our cluster sample, we found
that effects of interstellar reddening are far from negligible for the indices
covering the widest wavelength intervals, in particular Mg~Wide. Index
variations turned out to be of the order of 10\% for an E(B-V) difference of
0.1~mag.


\subsection{Globular Cluster Indices vs.\ Chemical Composition}
\label{sec:index_metal}

\begin{figure*}[!t]
\begin{center}
\includegraphics[height=16.0cm]{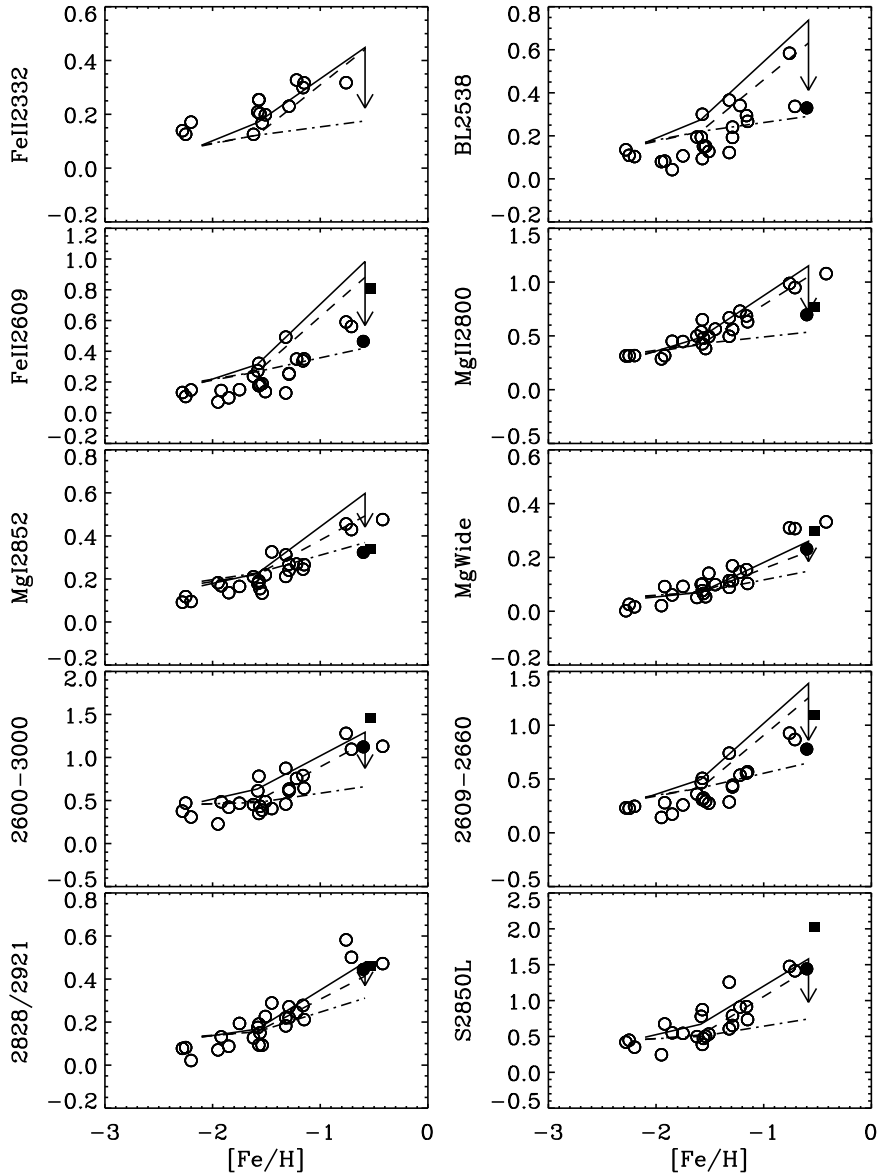}
\caption{Observed GGC selected indices
compared to synthetic indices of SSP of 10 (solid line), 12 (dashed line), and 13~Gyr 
(dot-dashed line).
Models are the same of Fig.~\ref{fig:seds_indexteffage}.
The solid circle shows the position of NGC~6388,
while the solid square represents
NGC~6441. The vertical arrows at the extrema of the 10Gyr synthetic indices show the 
dilution of an index due to the presence of hot stars in the HB (see section 4.2 
for details).
\label{ggcindex_vs_met}}
\end{center}
\end{figure*}

In Fig.~\ref{ggcindex_vs_met}, we show the trends of the ten selected indices
vs.\ metallicity for our sample of 27 GGCs. Among the main features are:

a)- All indices steadily increase with increasing chemical composition and
display remarkable good correlations, most of them increasing a factor of 5--6
with [Fe/H] increasing from about $-2.2$ to $-0.5$ dex. In the work of
\citet{rd99} the correlation was explored aimed at assessing the adequacy of
metal-rich clusters for studying the populations of M32. In addition to their
five indices, we found that five more also display a monotonic behavior up to
the metal-rich edge of our sample. From the analysis of synthetic populations,
we found that this monotonic tendency is, general, continued at super-solar regimes. An
exception to this trend is, for instance, the Mg~\textsc{ii}~2800), for which the oldest 
populations present an inflexion and indices decrease towards high metallicities.

b)-  It is interesting to note that, while the spread in age to some
extent increases the vertical dispersion of the points, age is not the main
agent, and the dispersion should be abscribed to other reasons, in particular
the UV bright population (stars at the turn-off, hot horizontal branch stars, and
blue stragglers). In this line, from an analysis of the Mg~\textsc{ii}~2800 index,
\citet{rd99} concluded that the observed metallicity-index correlation can be
explained by the dependence of the temperature of the TO on metallicity. They
also examined the effects of the HB morphology and found no correlation
between the dispersion and HB morphology. Such results are confirmed by
scrutiny of the panels in Fig.~\ref{ggcindex_vs_met} only for low metallicity systems, where no clear tendency
is evident. As we shall see in what follows, the presence of extended HBs severely affects some of the indices.

As a supplemental piece of the puzzle, it is noteworthy to mention that there
is evidence that enhancement of elements formed by capture of $\alpha$
particles is present in globular clusters, reflecting the overall chemical evolution of the Milky Way \citep{lw05,mendel07}. In principle, such effects should be taken into account in both the available stellar data sets that include spectra with $\alpha$-enhancement \citep[e.g.,][]{munari05} and in the opacities used to construct evolutionary tracks. Whilst the detailed study of the imprints of $\alpha$-element enrichment in mid-UV indices is beyond the scope of this paper, we consider it convenient to present in \S~\ref{alpha} a brief analysis.

\subsection{Synthetic vs.\ Empirical Indices}\label{glob_vs_synt}

In order to quantitatively compare empirical indices with results from
population synthesis techniques, we have initially considered to apply, whenever
possible, the  transformation coefficients needed to match theoretical and
empirical indices of stars as described in Paper II. However, in composite
energy distributions, the discussed limitations of the theoretical atmospheric
data are hauled together with additional potential drawbacks of their own. Instead,
we have conducted
a direct comparison of results of the synthesis codes after matching
the {\it IUE} nominal resolution in low dispersion mode
of 6~\AA.

In Fig.~\ref{ggcindex_vs_met}, we have superimposed to the data of globular
clusters the synthetic integrated indices for the ages of 10~Gyr (solid line), 12~Gyr (dashed line), and 13~Gyr
(dot-dashed line). Only indices for Z=0.0001, Z=0.0004, and
Z=0.004 are plotted here, because they span the range of the
observed values. It is worth stressing that models
are plotted in this diagram with the [Fe/H] values corresponding
to the metallicity of the isochrones, derived from the
abundances of the atmosphere models.
Finally, we remind that the solid circle represents the NGC~6388
and the solid square indicates NGC~6441.
These two clusters display quite extended horizontal branches as
compared to the other three metal rich clusters (characterized by the presence
of a red clump), indicating the possibility of a higher
than normal helium content \citep[e.g.,][]{bcf94}.

Synthetic indices fairly well reproduce the trends of the empirical
data. However, in some cases, especially for the lower metallicity clusters,
the models tend to overestimate the data. This is particularly evident in the
case of BL~2538, Fe~\textsc{ii}~2609, 2609/2660, and 2828/2921. We also notice
that age differences, at fixed metallicity, are visible only for the highest
metallicity bin.
In what follows, we give some remarks on groups of indices, while a
more global discussion is deferred to \S~\ref{problems}.

a)- Fe~\textsc{ii}~2332: The overall trend is faithfully reproduced by the
synthetic indices. This index is the only case where there is a slight
theoretical underestimation of about 0.05~mag.

b)- BL~2538, Fe~\textsc{ii}~2609, and 2609/2660: These indices show general
offsets, which are about 0.2~mag at the lower metallicities. However, at the
highest metallicity, almost all the data are well bracketed by the 10~Gyr and
13.5~Gyr models. In these indices, NGC~6388 is always consistent with a
large age, while NGC~6441 is compatible with different ages
depending on the adopted index.
This fact likely reflects the different stellar
mixture, in particular the presence of a more or less prominent
(with respect to the metallicity) ensemble of hot He-burning stars.

According to the correlations seen in stars, the differences between
theoretical and {\it IUE} indices (obtained 
from the least square fit presented in Paper II or estimated by
inspection of Fig.~6 in that paper), for the largest measured {\it IUE}
indices, are approximately 0.34, 0.5, and 0.5~mag. Therefore the overestimation
of these three indices can be amply explained by stellar discrepancies.

c)- Mg~\textsc{ii}~2800: The models match the data fairly well.
All clusters follow a line of constant age around 10--12~Gyr.
Interestingly, the clusters NGC~6388 and NGC~6441 are both well
separated from the other three metal-rich clusters,
reflecting the presence of the hot HB population.

d)- Mg~\textsc{i}~2852: It also shows a quite fair match, apart from the
lowest metallicity case, where the models sligthly overestimate the empirical
data. At high metallicity, the age differences are more pronounced than
in the case of Mg~\textsc{ii}~2800. This index indicates ages around 12~Gyr
for the high metallicity clusters, apart from the two outliers
already discussed above.

e)- Mg~Wide and 2828/2921: These are amongst the best reproduced indices, as
was the case for the stellar indices. It is important to
note that Mg~Wide displays the least dispersion, most probably reflecting the
dilution of discrepancies due to its broader bands. Paradoxically, the features
that are expected to modulate this index are not very well reproduced
individually.Note that for these two indices the clusters NGC~6388 and NGC~6441 are close to
those of the other clusters of similar metallicity, and
consistent with an age of about 12 Gyr.

f)- 2600-3000 and S2850L: These two continuum indices are formed by widely
separated bands. The match with the data is fairly good 
and, similarly to indices mentioned in the previous point, metal-rich
clusters display consistent indices, with NGC~6441 attaining the highest
values.

The behavior depicted by the indices in the last two points above is interesting and may be 
due to a certain independence of the indices from the extension
of the HB. This will be investigated in a forthcoming paper 
(see also section 4.2).

\section{Theoretical Stellar Data Bases: a Need for an Update} \label{problems}

\subsection{Shortcomings of the theoretical spectra} \label{bad_theory}
We have extensively discussed the caveats associated with the UVBLUE spectral
library in Paper~I and their implications in the synthetic stellar indices in
Paper~II. In order to decrease the observed discrepancies in stars, it would be
important to re-calculate the grid following a two step process 
(in addition to the calculation of the grid itself). We first
need to incorporate up-dated solar abundances, since the parent model
atmospheres of UVBLUE spectra adopted the chemical composition of
\citet{and_grev89}, while new Kurucz model atmospheres are now available with
the \citet{grev_sauv98} abundances. The differences in these two databases
result in a change of the global metallicity $Z$ of the order of 0.002. We
actually conducted a series of tests by comparing entries in UVBLUE with those
of \citet{munari05}, which incorporates \citet{grev_sauv98} abundances, and
the effects on the overall energy distribution are negligible. More recently,
however, 3-D hydrodynamic calculations by \citet{gas07} have shown that a
major update of solar composition might be required. These latest results
indicate a solar metallicity about one half ($Z=0.012$) of the value reported
in the papers cited above. It is therefore of fundamental importance that
these new values are tested and included in the synthetic atmospheres and
interiors. Once the reference abundances have been anchored, the second step,
perhaps as important as the first one, consists on the evaluation of the
atomic line parameters. The fact that theoretical spectra predict too deep
absorption lines \citep{bell94,bertone08} can be, in some cases, solely
ascribed to the uncertain line parameters (mainly oscillator strengths and Van
der Waals damping constants). A comparison with high-quality and
high-resolution observational data, as in \citet{peterson05}, will certainly
bring the solution to this problem.

\subsection{Theoretical Incompleteness of the Hot Evolutionary Stages}

We have mentioned that hot stars play an important role in modulating
(decreasing) the index strength. Evolutionary tracks should, in principle,
include evolutionary prescriptions that allow the formation of objects that
populate extensions of the main sequence and the horizontal branch in the
color-magnitude diagram of stellar clusters (such as BSs,
blue HB stars, and components of the so-called blue tail and
blue hook of the HB). As far as BSs are concerned, it has been demostrated that
their presence in old open clusters severely affects the integrated
energy distributions, particularly at ultraviolet wavelengths
\citep{Xin07}. More pronounced effects are expected from He-burning objects
that attain high luminosities and temperatures (higher than those expected
from metallicity effects alone) as those present in some GGCs
\citep[see][for the cases of NGC~6441 and NGC~6388]{busso07}. It is clear that a
more appropiate way to carry out the comparisons presented in
\S~\ref{glob_vs_synt} would preferably have to include in the synthesis code
the observed star counts and distributions (extracted from CM diagrams) of
these hot components for each cluster. However, even if we are able to include
the effects of these hot stars, such effects  should be modelled on the basis
of an integrated property of the population in order to be applicable to the
analysis of non-resolved stellar aggregates, which is the main scope of this
paper. Perhaps an interesting suggestion has emerged from the magnesium
indices depicted in Fig.~\ref{ggcindex_vs_met}, which appear to segregate
clusters with blue HBs.

\begin{figure}[!t]
\begin{center}
\includegraphics[height=6.cm,width=7.0cm]{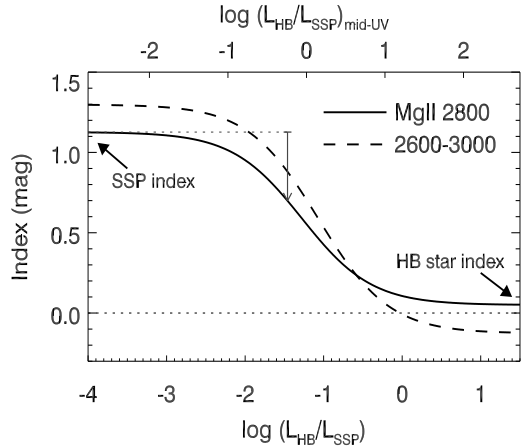}
\caption{Dilution of the Mg~\textsc{ii}~2800 and 2600-3000 indices as a function of the bolometric luminosity 
fraction of artificially added hot stars with respect to the total luminosity of the original population: $L_{\rm
HB}/L_{\rm SSP}$. The thin vertical arrow stands for the index decrease that would correspond to the difference between
the averaged empirical values of the clusters NGC~6388 and NGC~6441 and that of an 10~Gyr population, indicating that
a luminosity ratio of $log(L_{\rm
HB}/L_{\rm SSP})$ = -1.46 can explain that difference. By fixing this ratio we obtained similar {\it dilution} vectors 
for the rest of the indices. In the x-axis at the top we include the scale for the integrated flux ratio in the mid-UV.
\label{hot_stars}}
\end{center}
\end{figure}

In Fig.~\ref{hot_stars}, we show the results of a simple but illustrative
exercise in which we analyze the effects of the presence of hot stars on two
of the indices included in Fig.~\ref{ggcindex_vs_met}. For this purpose, we
have artificially added a hot stellar component to a set of synthetic
populations of 10~Gyr and $Z$=0.004.  The stellar model flux that we have
added in different amounts, corresponds to the parameters of the hottest HB
star present in the stellar isochrones of Fig.~\ref{ggcindex_vs_met} (Age/$T_{\rm
eff}$/$\log{g}$/[Fe/H])=(13~Gyr/10490/3.72/$-0.68$). At any rate, the qualitative behavior of the results 
does not depend of the exact value of the
stellar parameters. In the figure, we plot the index
strength as a function of the bolometric luminosity fraction of the added hot
stars with respect to the total luminosity of the original population: $L_{\rm
HB}/L_{\rm SSP}$. The figure shows that indices preserve their starting value
(that of the SSP without any additional hot star) almost constant until the HB
luminosity accounts for approximately 1\% of that of the parent population. At
about $L_{\rm HB}=L_{\rm SSP}$ the trends asymptotically approach the values
of the HB stars alone. It is important to remark that we have actually
computed the HB effects for the remaining eight indices and they all display a
quite similar behavior. This analysis indicates, as anticipated, that the 
presence of hot objects in a stellar population is reflected in a reduction of their
mid-UV indices, and, consequently, could provide an explanation for
the low values of the line indices, in particular for the cluster NGC~6441.
 
In order to quantitatively establish a {\it dilution} vector for each index,
we have taken the index Mg~\textsc{ii}~2800 as a reference. We assumed that the 
low values of this index in NGC~6388 and NGC~6441 are solely due to hot HB stars.
Then, we have calculated the difference between
the averaged empirical values of the clusters NGC~6388 and NGC~6441 and that of 
an 10~Gyr population (0.42mag, indicated with an vertical arrow in 
Fig.~\ref{hot_stars}) and searched for the 
luminosity fraction that accounts for this difference; $log (L_{\rm HB}/L_{SSP}) = -1.46$.
With this luminosity ratio we obtained the vectors for the rest of the indices and 
are indicated with arrows in Fig.~\ref{ggcindex_vs_met} on the right of each panel. Note that
these {\it dilution} vectors allow also an explanation for the low index values
for BL~2538, Fe~\textsc{ii}~2609, Mg~\textsc{i}~2852, and 2609/2660 of the cluster
NGC~6388 and for Mg~\textsc{i}~2852 of NGC~6441. To within the uncertainties
associated with the empirical indices Mg~Wide and 2828/2921 and considering
the their rather small vectors, these last two indices are also
appropriately reproduced for both clusters.

The above analysis, however, demands an interpretation of the inconsistent
values  of NGC~6441 for the indices Fe~\textsc{ii}~2609, 2609/2660 and
S2850L. We have double checked the only available IUE image of NGC~6441 and found
that the flux errors associated with the blue bands defining these indices
translate into significantly larger uncertainties on the indices of 
NGC~6441 than those of NGC~6388. 

Somewhat intriguing is the behavior of the 2600--3000 slope index. The two
metal-rich clusters with hot HBs are about in the same loci as the red HB
clusters, eventhough the above analysis indicates that the vector for this
index is large (0.42~mag).  One can speculate on a number of reasons that potentially 
affect the overall mid-UV energy distribution (such as color excess), nevertheless, 
from an empirical point of view it appears that this index is, to some extent, insensitive to the 
dilution by hot stars. We want to stress that a more profound examination is
required. This has to include, as mentioned earlier, the appropiate numbers
and distributions of the hot stellar component. Nevertheless, it is clear that
the exclusion of hot objects could lead to wrong predictions for metal rich
systems.

\section{Effects of Alpha Enhancement on the Mid-UV Morphology}\label{alpha}

Closely related to the brief discussion in \S~\ref{bad_theory} is the evidence that the elemental abundances in globular clusters show significant variations with respect to solar partition values. It has been a common practice to incorporate, within population synthesis techniques, adjustments to optical indices (in both theoretical and empirical) that account, for example, for the enhancement of the $\alpha$-elements (O, Na, Ti, Ca, Mg, etc.) \citep{mendel07}. In the mid-UV, the effects of $\alpha$-enhancement have never been explored and in this section we present some preliminary results.

In what follows, we made use of the theoretical database of stellar fluxes
available at Fiorella Castelli's
website\footnote{http://wwwuser.oat.ts.astro.it/castelli/grids/}
\citep{castelli03}. In view of the low resolution of these grids and within
the exploratory nature of this investigation, we decided to concentrate on
showing the effects of $\alpha$-enhancement on the theoretical stellar and
population SEDs, defering the detailed analysis of the impact of these effects
on continuum and line indices for a future investigation in which we also partially cover the points mentioned in the previous section. An important note is that the grid of Castelli's stellar fluxes has been calculated by adopting the \citet{grev_sauv98} solar abundances and consider a set of updated opacity distribution functions (NEWODFs).

\begin{figure*}[!t]
\begin{center}
\includegraphics[height=16.cm,width=16.0cm]{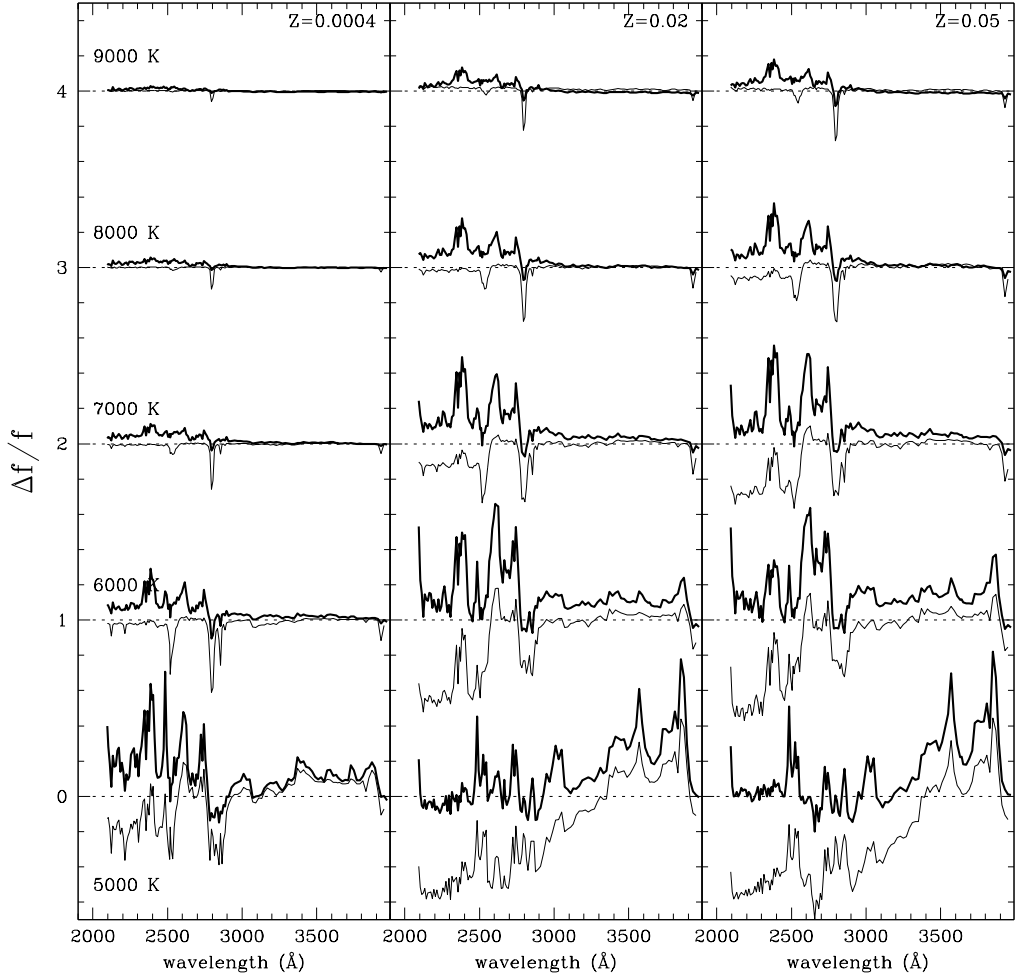}
\caption{Effects of $\alpha$-enhancement in stellar theoretical fluxes for a
  fixed surface gravity of $\log{g}$=4.0~dex and different metallicities and
  effective temperatures.  The y-axis depicts the relative flux ratio in the
  form of $\Delta f/f$ where $\Delta f=f_{\alpha}-f$ is the difference between
  the flux with an $\alpha$-enhancement of [$\alpha$/Fe]=+0.4~dex and those
  considering [$\alpha$/Fe]=+0.0~dex. For the sake of clarity the flux ratios
  have been scaled, from bottom to top, with a unity offset. 
The thick solid line shows the case where the $\alpha$-enhancement is
accompanied by a rescaling of all metals to preserve the global metallicity,
while the thin solid line refers to the case where this rescaling is not
performed (see text for more details). 
\label{fig:star_alpha}}
\end{center}
\end{figure*}

In Fig.~\ref{fig:star_alpha}, we show the residuals for a series of
theoretical fluxes for three chemical compositions ($Z$ = 0.0004, 0.02, 0.05),
five different temperatures, and a surface gravity of $\log{g}$=4.0~dex. In the
lower left corner in each panel on the left we indicate the effective
temperature for that row. The vertical axes are the flux residuals $\Delta
f/f$, where $\Delta f=f_{\alpha}-f$ is the difference between the flux with an
$\alpha$-enhancement of [$\alpha$/Fe]=+0.4~dex and those considering
[$\alpha$/Fe]=+0.0~dex. The residuals for temperatures $T_{\rm eff} > 5000$~K have been scaled with offsets of 1, 2, 3, and 4 with respect to that for 5000~K. The thick lines illustrate the relative residuals for the cases in which $Z$ has been fixed. A significant
$\alpha$-enhancement, keeping the total metallicity $Z$ fixed,
implies that the abundances of the non-enhanced
metals (in particular Fe)
have to be rescaled downwards \citep[see, for example,][for the case of Lick indices]
{annibali07,mendel07}. Since oxygen is the most abundant metal,
the decrease on the non-$\alpha$ metals turns out to be significant.
Particularly important for the stellar atmospheres is the decrease of Fe
because of its large contribution to line blanketing.
As an example, in the case of $Z=0.004$, the atmosphere models have
[Fe/H]=$-0.58$ for non-enhanced abundances and [Fe/H]=$-0.89$ if
[$\alpha$/Fe]=+0.4~dex.
To perform this rescaling we have simply interpolated
the needed stellar atmospheres.

Note that the residuals heavily depend on temperature and chemical
composition, with the $\alpha$-enhanced fluxes being on average larger in the
mid-UV and hence delivering positive residuals. The residuals can reach values
in excess of 60\%, particularly at the solar and super solar regimes.

The thin lines in Fig.~\ref{fig:star_alpha} show the residuals when we compare
theoretical fluxes with the same nominal parameters ($T_{\rm
  eff}$/$\log{g}$/[M/H]) of the parent model, i.e.\ $Z$ is not fixed. In this case the 
residuals are generally negative, reflecting the overestimation of the blanketing.

For the calculation of the composite $\alpha$-enhanced spectra of SSPs, we
remind that the new Padova models do not account yet for
variations of $\alpha$ elements partition.
Thus, in the calculation of present SSP spectra,
the $\alpha$-enhancement is only taken into account for
the effect it has in the stellar atmospheres. The stellar fluxes used in
the synthesis code were those with a renormalized heavy element abundance to
a fixed metallicity ($Z$). The effects on the stellar evolutionary tracks
should not be significant for the low metallicities of GGCs.

\begin{figure*}[!t]
\begin{center}
\includegraphics[height=14.cm,width=16.0cm]{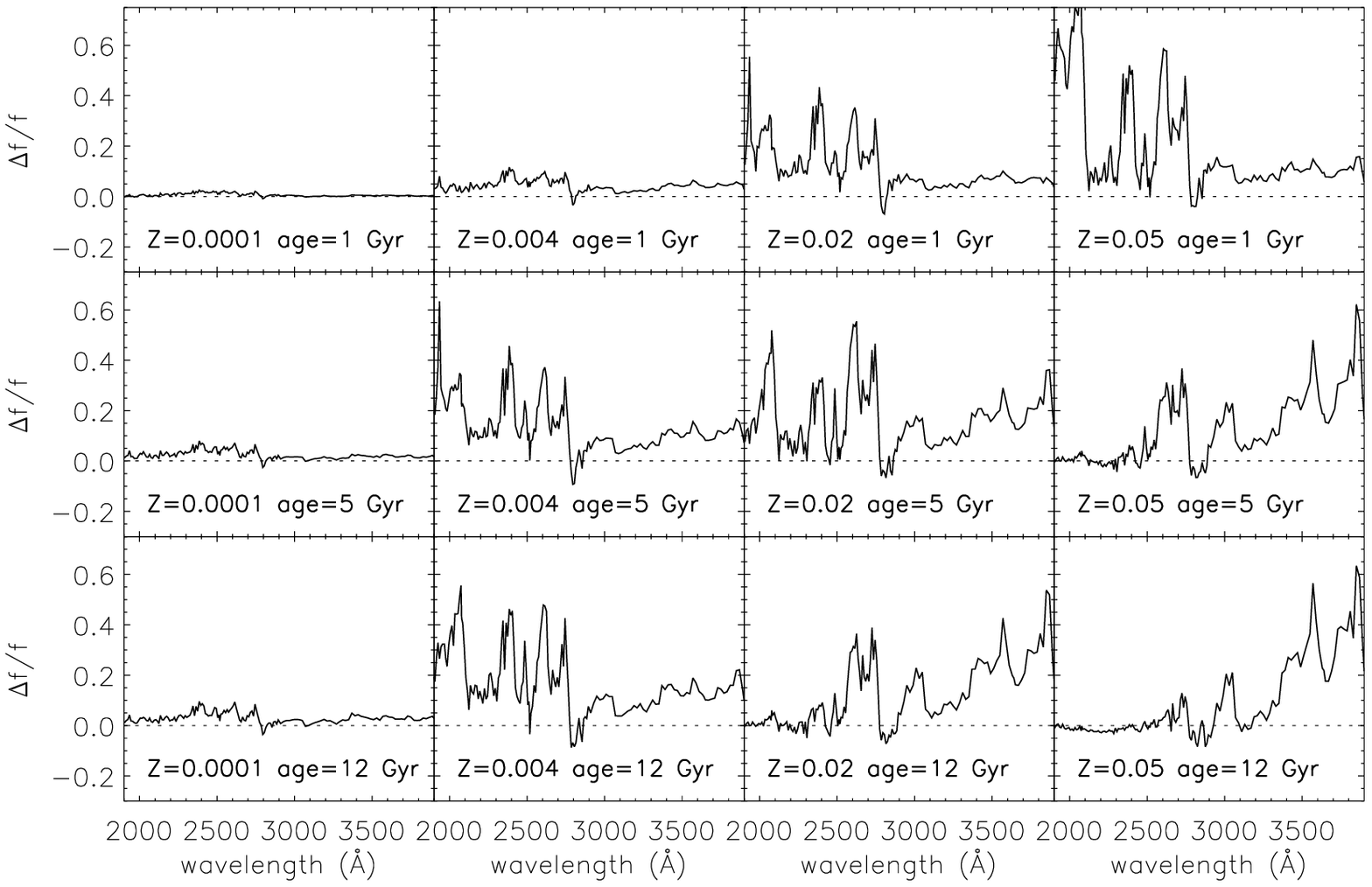}
\caption{Effects of $\alpha$-enhancement of SSP models for different ages (rows)
and chemical compositions (columns). The y-axis illustrates the flux ratio as in Fig.~\ref{fig:star_alpha}. In this comparison global metallicity of
the $\alpha$-enhanced stellar fluxes have been properly rescaled to preserve the same $Z$.
\label{fig:ssp_alpha}}
\end{center}
\end{figure*}

In Fig.~\ref{fig:ssp_alpha},  we show the relative differences between
integrated spectra of SSP with and without accounting for $\alpha$-enhancement.
The panels refer to the combinations of three ages (rows) and four chemical
compositions (columns). An examination of the figure allows to point out some
remarkable behaviors. First, the effects of $\alpha$-enhancement heavily
depend on $Z$, with the most pronounced differences at the metal-rich end,
even in the relatively young populations of 1~Gyr. For the most metal-poor
SSPs ($Z= 0.0001$), the flux differences never exceed 10\%. Second, concerning
the age effects, the notable differences seen in the mid-UV residual spectrum
of the 1~Gyr metal-rich population appear gradually shifted to longer
wavelengths. Globally, $\alpha$-enhancement is notoriously more important in
the mid-UV than in the near-UV, at ages and abundances compatible with those
of GGCs, and certainly far more important than in the optical, where it has
been extensively studied. This UV sensitivity suggests, among many other
possible options, to conduct an analysis of mid-UV indices of intermediate and
late-type stars in the context of the chemical evolution scenarios for the
Milky Way.

\section{Summary}\label{summary}
In this work we have studied the ultraviolet properties of evolved stellar populations,
with the purpose of analyzing the behavior of mid-UV indices in terms of
the leading population parameters (chemical composition and age) and checking
the current status of the theoretical models. With these results we trace the
next more important advances to perform.
The results can be summarized as follows:
\begin{itemize}
\item This is, to our knowledge, the first theoretical analysis of integrated
  mid-UV indices of old populations. They display a variety of behaviors,
  with the main one being the low index insensitivity to age in a wide age interval (age $>$ 2~Gyr). Two indices, Fe~\textsc{ii}~2332 and Fe~\textsc{ii}~2402, show, however, a remarkable distinct behavior with respect to other indices at the metal rich regime. This result is of particular importance, since our work is aimed at providing predictive tools for the analysis of elliptical galaxies of high metallicity.

\item We also present a test of the ability of SSP fluxes to reproduce 10
  spectroscopic 
  indices measured in the {\it IUE} spectra of a sample of 27 GGCs. Theoretical indices well
  reproduce the overall trends but show slight discrepancies in which synthetic
  indices are sistematically higher. In this respect, we conclude that SSPs
  well represent the observational prototypes of simple populations, after the
  appropiate scaling of SSPs synthetic indices to the {\it IUE}
  system. Therefore, one can be confident that
  the approach presented here serves as the basis for studies of metal-rich
  populations, which cannot be modelled with existing empirical data bases, due
  to their poor coverage of the metallicity space.

\item We provide a brief diagnostics on how the theoretical ingredients should
  be improved. We suggest the following three main roads: a)- improvement of the solar reference abundances in the calculation of
  theoretical atmospheres and evolutionary tracks; b)- semi-empirically
  up-dating of the atomic line parameters to reduce the too strong absorption
  lines in synthetic spectra; and c)- check with the new isochrones the
  importance of hot evolved evolutionary phases.

\item Together with the above points, it is important to incorporate
the effects of $\alpha$-enhancement in adequate resolution stellar atmosphere
models. In fact, the tests we have illustrated here, though performed with
low resolution models, have shown that an enhancement of +0.4 dex
may result in dramatic differences for the integrated spectra
of the same metallicity and age. In parallel, $\alpha$-enhancement should also be considered
for the computation of the evolutionary tracks, though this effect is expected
to be important only at high metallicity.
\end{itemize}

All the above issues will be the subjects of our future
investigations.

\acknowledgements M.C. and E.B. are pleased to thank financial support from
Mexican CONACyT, via grants 49231-E and SEP-2004-C01-47904. A.B. acknowledges
contract ASI-INAF COFIS I/016/07/0.


\clearpage

\begin{deluxetable}{lcrrcrcc}
\tablecaption{Main parameters of Globulars \label{glob_data}}
\tablewidth{0pt}
\tablehead{
\colhead{Cluster} & \colhead{[Fe/H]}    & \colhead{age}       &\colhead{$v_{\rm rad}$}& \colhead{E(B-V)}      & \colhead{HBR}      & \colhead{hbt}  & \colhead{notes} \\
        &\colhead{(dex)}     & \colhead{(Gyr)}     &\colhead{(km~s$^{-1}$)}  &  \colhead{(mag)}      &          &      &
}
\startdata
NGC~104   &  -0.76  &     10.7  &     -18.7  &       0.04  &   -0.99  &  7  &       \\
NGC~362   &  -1.16  &      8.7  &     223.5  &       0.05  &   -0.87  &  6  &       \\
NGC~1851  &  -1.22  &      9.2  &     320.5  &       0.02  &   -0.36  &  1  &       \\
NGC~1904  &  -1.57  &     11.7  &     207.5  &       0.01  &    0.89  &  0  &       \\
NGC~2808  &  -1.15  &      9.3  &      99.7  &       0.22  &   -0.49  &  0  &       \\
NGC~5272  &  -1.57  &     11.3  &    -147.1  &       0.01  &    0.08  &  4  &       \\
NGC~5824  &  -1.85  &  9.2-10.4 &     -27.5  &       0.13  &    0.79  &  2  &  (a)  \\
NGC~5904  &  -1.29  &     10.9  &      51.8  &       0.03  &    0.31  &  3  &       \\
NGC~6093  &  -1.75  &     12.4  &       7.3  &       0.18  &    0.93  &  0  &       \\
NGC~6205  &  -1.54  &     11.9  &    -246.6  &       0.02  &    0.97  &  0  &       \\
NGC~6266  &  -1.29  &  11.0-12.1&     -68.0  &       0.47  &    0.32  &  4  &  (a)  \\
NGC~6284  &  -1.32  &     11.0  &      29.7  &       0.28  &    9.99  &  1  &       \\
NGC~6293  &  -1.92  &   7.9-8.8 &     -98.9  &       0.41  &    0.90  &  1  &  (a)  \\
NGC~6341  &  -2.28  &     12.3  &    -120.5  &       0.02  &    0.91  &  2  &       \\
NGC~6388  &  -0.60  &     10.7  &      81.2  &       0.37  &    9.99  &  5  &  (b)  \\
NGC~6397  &  -1.95  &     12.1  &      18.9  &       0.18  &    0.98  &  1  &       \\
NGC~6441  &  -0.53  &     11.0  &      18.3  &       0.47  &    9.99  &  7  &  (c)  \\
NGC~6624  &  -0.42  &     10.6  &      54.3  &       0.28  &   -1.00  &  7  &       \\
NGC~6626  &  -1.45  &     12.6  &      17.0  &       0.40  &    0.90  &  2  &  (d)  \\
NGC~6637  &  -0.71  &     10.6  &      39.1  &       0.16  &   -1.00  &  7  &       \\
NGC~6681  &  -1.51  &     11.5  &     218.7  &       0.07  &    0.96  &  1  &       \\
NGC~6715  &  -1.58  &     12.2  &     141.9  &       0.15  &    0.75  &  3  &  (e)  \\
NGC~6752  &  -1.56  &     12.2  &     -27.4  &       0.04  &    1.00  &  0  &       \\
NGC~6864  &  -1.32  &      9.0  &    -189.3  &       0.16  &   -0.07  &  1  &  (a)  \\
NGC~7078  &  -2.25  &     11.7  &    -106.6  &       0.10  &    0.67  &  0  &       \\
NGC~7089  &  -1.62  & 11.8-13.7 &      -6.7  &       0.06  &    0.96  &  1  &  (a)  \\
NGC~7099  &  -2.20  &     11.9  &    -184.3  &       0.03  &    0.89  &  1  &       \\
\enddata
\tablecomments{(a) Ages from Case (A) of \citet{mw06}. (b) We adopted as the age of NGC~6388 that
of NGC~104 according to \citet{catelan06}. (c) \citet{cd07} assumed a reference age of 11.0~Gyr.
(d) \citet{testa01} indicate that NGC~6626 is coeval with respect to NGC~2298 and this latter cluster has 12.6~Gyr in \citet{sw02}. (e) Age from \citet{rakos05} derived from Str\"{o}mgren photometry.}
\end{deluxetable}

\clearpage

\begin{deluxetable}{lcccccccccc}
\tabletypesize{\scriptsize}
\rotate
\tablecaption{Mid-UV indices for the sample clusters \label{glob_index}}
\tablewidth{0pt}
\tablehead{
\colhead{object} & \colhead{Fe~\textsc{ii}~2332} & \colhead{BL~2538} & \colhead{Fe~\textsc{ii}~2609} & \colhead{Mg~\textsc{ii}~2800} &
\colhead{Mg~\textsc{i}~2852} & \colhead{Mg~Wide} & \colhead{2600-3000} &
\colhead{2609/2660} & \colhead{2828/2921} & \colhead{S2850L}
}
\startdata
NGC~104     & 0.32    & 0.58    & 0.59    & 0.99    & 0.46    & 0.31    & 1.28    & 0.93    & 0.58    & 1.48 \\
NGC~362     & 0.30    & 0.29    & 0.34    & 0.69    & 0.25    & 0.15    & 0.79    & 0.56    & 0.28    & 0.91 \\
NGC~1851    & 0.33    & 0.34    & 0.35    & 0.73    & 0.27    & 0.15    & 0.76    & 0.54    & 0.25    & 0.91 \\
NGC~1904    & 0.25    & 0.09    & 0.18    & 0.43    & 0.19    & 0.08    & 0.35    & 0.31    & 0.09    & 0.39 \\
NGC~2808    & 0.32    & 0.27    & 0.35    & 0.63    & 0.27    & 0.10    & 0.64    & 0.57    & 0.21    & 0.73 \\
NGC~5272    & 0.25    & 0.30    & 0.32    & 0.65    & 0.19    & 0.10    & 0.78    & 0.51    & 0.19    & 0.87 \\
NGC~5824    & \nodata & 0.04    & 0.10    & 0.45    & 0.14    & 0.06    & 0.42    & 0.17    & 0.09    & 0.55 \\
NGC~5904    & 0.23    & 0.19    & 0.25    & 0.56    & 0.27    & 0.11    & 0.63    & 0.44    & 0.22    & 0.65 \\
NGC~6093    & \nodata & 0.11    & 0.15    & 0.45    & 0.17    & 0.09    & 0.47    & 0.26    & 0.19    & 0.54 \\
NGC~6205    & 0.17    & 0.15    & 0.19    & 0.38    & 0.13    & 0.05    & 0.39    & 0.29    & 0.09    & 0.50 \\
NGC~6266    & \nodata & 0.24    & 0.25    & 0.56    & 0.24    & 0.17    & 0.61    & 0.43    & 0.27    & 0.79 \\
NGC~6284    & \nodata & 0.12    & 0.13    & 0.50    & 0.21    & 0.09    & 0.46    & 0.29    & 0.18    & 0.61 \\
NGC~6293    & \nodata & 0.08    & 0.14    & 0.32    & 0.17    & 0.09    & 0.48    & 0.28    & 0.13    & 0.67 \\
NGC~6341    & 0.14    & 0.14    & 0.13    & 0.31    & 0.09    & 0.00    & 0.38    & 0.23    & 0.08    & 0.42 \\
NGC~6388    & \nodata & 0.33    & 0.46    & 0.70    & 0.32    & 0.23    & 1.12    & 0.78    & 0.44    & 1.44 \\
NGC~6397    & \nodata & 0.08    & 0.07    & 0.29    & 0.18    & 0.02    & 0.23    & 0.14    & 0.07    & 0.24 \\
NGC~6441    & \nodata & \nodata & 0.81    & 0.77    & 0.34    & 0.30    & 1.46    & 1.09    & 0.46    & 2.03 \\
NGC~6624    & \nodata & \nodata & \nodata & 1.08    & 0.48    & 0.33    & 1.13    & \nodata & 0.47    & \nodata \\
NGC~6626    & \nodata & \nodata & \nodata & 0.56    & 0.33    & 0.10    & 0.41    & \nodata & 0.29    & \nodata \\
NGC~6637    & \nodata & 0.34    & 0.56    & 0.95    & 0.43    & 0.31    & 1.10    & 0.87    & 0.50    & 1.41 \\
NGC~6681    & 0.20    & 0.13    & 0.14    & 0.49    & 0.22    & 0.14    & 0.49    & 0.28    & 0.23    & 0.53 \\
NGC~6715    & 0.21    & 0.19    & 0.27    & 0.54    & 0.18    & 0.10    & 0.61    & 0.46    & 0.17    & 0.78 \\
NGC~6752    & 0.20    & 0.15    & 0.18    & 0.48    & 0.16    & 0.07    & 0.43    & 0.32    & 0.15    & 0.47 \\
NGC~6864    & \nodata & 0.37    & 0.49    & 0.67    & 0.31    & 0.11    & 0.87    & 0.74    & 0.22    & 1.26 \\
NGC~7078    & 0.13    & 0.11    & 0.11    & 0.32    & 0.12    & 0.03    & 0.47    & 0.23    & 0.08    & 0.45 \\
NGC~7089    & 0.13    & 0.19    & 0.24    & 0.50    & 0.21    & 0.05    & 0.45    & 0.36    & 0.13    & 0.50 \\
NGC~7099    & 0.17    & 0.10    & 0.15    & 0.32    & 0.10    & 0.02    & 0.31    & 0.25    & 0.02    & 0.35 \\
\enddata
\end{deluxetable}


\begin{thebibliography}{99}

\bibitem[Anders \& Grevesse(1989)]{and_grev89} Anders, E., \& Grevesse, N.\ 1989, Geochim. Cosmochim. Acta, 53, 197

\bibitem[Annibali et al.(2007)]{annibali07} Annibali, F., Bressan, A., Rampazzo, R., Zeilinger, W.~W., \& Danese, L.\ 2007, \aap, 463, 455

\bibitem[Bell et al.(1994)]{bell94} Bell, R.~A., Paltoglou,
G., \& Tripicco, M.~J.\ 1994, \mnras, 268, 771

\bibitem[Bertelli et al.(1994)]{bertelli94} Bertelli, G., Bressan, A., Chiosi,
  C., Fagotto, F., \& Nasi, E.\ 1994, \aap, 106, 275

\bibitem[Bertelli et al.(2008)]{bertelli08} Bertelli, G., Girardi, L., Marigo, P., \& Nasi, E.\ 2008, \aap, 484, 815

\bibitem[Bertola et al.(1995)]{bertola95} Bertola, F., Bressan,
A., Burstein, D., Buson, L.~M., Chiosi, C.,
\& di Serego Alighieri, S.\ 1995, \apj, 438, 680

\bibitem[Bertone et al.(2007)]{bertone07} Bertone, E., Buzzoni, A. Chavez, M.,
\& Rodriguez-Merino, L. H.\ 2007, in  ASP Conf. Ser. 374, From Stars to
Galaxies: Building the pieces to build up the Universe, ed. A. Vallenari,
R. Tantalo, L. Portinari, \& A. Moretti (San Francisco, CA: ASP) p.~399

\bibitem[Bertone et al.(2008)]{bertone08} Bertone, E., Buzzoni, A., Chavez, M., \& Rodriguez-Merino, L.~H.\ 2008, \aap, 485, 823

\bibitem[Bonatto et al.(1995)]{bba95} Bonatto, C., Bica, A., \& Alloin, D.\ 1995, \aap, 112, 71

\bibitem[Bressan et al.(1994)]{bcf94} Bressan, A., Chiosi, C., \& Fagotto, F.\
1994, \apjs, 94, 63

\bibitem[Bressan et al.(1998)]{bgs98} Bressan, A., Granato, G. \& Silva, L.\ 1998,
\aap, 332, 135

\bibitem[Brown et al.(2008)]{brown08} Brown, T.~M., Smith, E.,
Ferguson, H.~C., Sweigart, A.~V., Kimble, R.~A.,
\& Bowers, C.~W.\ 2008, \apj, 682, 319

\bibitem[Busso et al.(2007)]{busso07} Busso, G., et al.\ 2007, \aap, 474, 105

\bibitem[Buzzoni et al.(2009)]{buzzoni08} Buzzoni, A., Bertone, E., Chavez, M. \& Rodriguez-Merino, L. H.\ 2008,
in New Quests in Stellar Astrophysics II: The Ultraviolet Properties of
Evolved Stellar Populations, eds. M. Chavez, E. Bertone, L. H. Rodriguez-Merino
\& D. Rosa-Gonzalez (New York, NY: Springer), p.~263 (arXiv:0709.2711)

\bibitem[Caloi \& D'Antona(2007)]{cd07} Caloi, V., \& D'Antona, F. D.,\ 2007,
  \aap, 463, 949

\bibitem[Cardelli et al.(1989)]{ccm89} Cardelli, J. A., Clayton, G. C., \&
  Mathis, J. S.\ 1989, \apj, 345, 245

\bibitem[Carraro et al.(1996)]{carraro96} Carraro, G., Girardi, L., Bressan, A., \& Chiosi, C.\ 1996, \aap, 305, 849

\bibitem[Catelan et al.(2006)]{catelan06} Catelan, M., Stetson, P.~B., Pritzl,
  B. J., Smith, H.~A., Kinemuchi, K., Layden, A.~C., Sweigart, A.~V., \& Rich, R. M.\ 2006, \apjl, 651, 133

\bibitem[Castelli \& Kurucz(2003)]{castelli03}Castelli, F., \& Kurucz, R.~L.\
  2003, in IAU Symp. 210, Modelling of Stellar Atmospheres, ed. N. Piskunov,
  W.~W. Weiss \& D.~F. Gray (San Francisco, CA: ASP), A20

\bibitem[Chavez et al.(2007)]{chavez07} Chavez, M., Bertone, E., Buzzoni,
  A., Franchini, M., Malagnini, M.~L., Morossi, C., \& Rodriguez-Merino,
  L.~H.\ 2007, \apj, 657, 1046 (Paper~II)

\bibitem[Chavez(2009)]{chavez08} Ch{\'a}vez, M.\ 2009, \apss, 320, 45 

\bibitem[Clemens et al.(2009)]{clemens09} Clemens, M.~S.,
Bressan, A., Panuzzo, P., Rampazzo, R., Silva, L., Buson, L.,
\& Granato, G.~L.\ 2009, \mnras, 392, 982

\bibitem[Dickens(1972)]{dickens72} Dickens, R. F.\ 1972, \mnras, 157, 281

\bibitem[Fanelli et al.(1990)]{fanelli90}
Fanelli, M.~N., O'Connell, R.~W., Burstein, D., \& Wu, C.\ 1990, \apj, 364,
272

\bibitem[Fanelli et al.(1992)]{fanelli92}
Fanelli, M.~N., O'Connell, R.~W., Burstein, D., \& Wu, C.\ 1992, \apjs, 82,
197

\bibitem[G\'omez de Castro et al.(2009)]{ana08} G\'omez de Castro, A. I., et
  al.\ 2008, in New Quests in Stellar Astrophysics II: The Ultraviolet
  Properties of Evolved Stellar Populations, ed. M. Chavez, E. Bertone,
  L. H. Rodriguez-Merino \& D. Rosa-Gonzalez (New York, NY: Springer), p.~319

\bibitem[Grevesse \& Sauval(1998)]{grev_sauv98} Grevesse, N., \& Sauval,
  A. J.\ 1998, Space. Sci. Rev., 85, 161

\bibitem[Grevesse et al.(2007)]{gas07} Grevesse, N., Asplund, M., \& Sauval, A. J.\ 2007, Space Sci. Rev., 130, 105

\bibitem[Harris(1996)]{harris96} Harris, W.~E.\ 1996, \aj, 112, 1487

\bibitem[Kurucz(1993)]{Kurucz93}
        Kurucz, R. L.\ 1993, CD-ROM No.~13, ATLAS9 Stellar Atmosphere
  Programs and 2 km/s Grid

\bibitem[Lee(1990)]{lee90} Lee, Y. W.\ 1990, \apj, 363, 159

\bibitem[Lee \& Worthey(2005)]{lw05} Lee, H., \& Worthey, G.\ 2005, \apjs, 160, 176

\bibitem[Lotz et al.(2000)]{lotz00} Lotz, J.~M.,
Ferguson, H.~C., \& Bohlin, R.~C.\ 2000, \apj, 532, 830

\bibitem[Maraston et al.(2009)]{maraston08} Maraston, C., Nieves Colmen{\'a}rez, L., Bender, R., \& Thomas, D.\ 2009, \aap, 493, 425 

\bibitem[Marigo et al.(2008)]{marigo08} Marigo, P., Girardi, L., Bressan, A., Groenewegen, M.~A.~T., Silva, L., \& Granato, G.~L.\ 2008, \aap, 482, 883

\bibitem[Meissner \& Weiss(2006)]{mw06} Meissner, F., \& Weiss, A.\ 2006, \aap, 456, 1085

\bibitem[Mendel et al.(2007)]{mendel07} Mendel, J.~T., Proctor, R.~N., \& Forbes, D.~A.\ 2007, \mnras, 379, 1618

\bibitem[Munari et al.(2005)]{munari05} Munari, U., Sordo, R., Castelli, F.,
  \& Zwitter, T.\ 2005, \aap, 442, 1127

\bibitem[Peterson et al.(2005)]{peterson05} Peterson, R. C., et al.\ 2005, in
  Astrophysics \& Space Science Library, Vol. 329, Starbursts: From 30 Doradus
  to Lyman Break Galaxies, ed. R. de Grijs \& R.~M. Gonzalez Delgado
  (Dordrecht: Springer), p.~61

\bibitem[Ponder et al.(1998)]{ponder98} Ponder, J. M., et al.\ 1998, \aj, 116, 2297

\bibitem[Rakos \& Schombert(2005)]{rakos05} Rakos, K., \& Schombert, J.\ 2005, \pasp, 117, 245

\bibitem[Rodr{\'{\i}}guez-Merino(2004)]{lino04}
Rodr{\'{\i}}guez-Merino, L.~H.\ 2004, PhD Thesis, INAOE, Mexico

\bibitem[Rodr{\'{\i}}guez-Merino et al.(2005)]{lino05}
Rodr{\'{\i}}guez-Merino, L.~H., Ch{\'a}vez, M., Bertone, E., \& Buzzoni, A.\
2005, \apj, 626, 411 (Paper~I)

\bibitem[Rose \& Deng(1999)]{rd99} Rose, J., \& Deng, S.\ 1999, \aj, 117, 2213

\bibitem[Salaris \& Weiss(2002)]{sw02} Salaris, M., \& Weiss, A.\ 2002, \aap,
388, 492

\bibitem[Schiavon(2007)]{schiavon07} Schiavon, R.\ 2007, \apjs, 171, 146

\bibitem[Spinrad et al.(1997)]{Spinrad97} Spinrad, H., Dey, A.,
Stern, D., Dunlop, J., Peacock, J., Jimenez, R., \& Windhorst, R.\ 1997,
\apj, 484, 581

\bibitem[Sollima et al.(2005)]{sollima05} Sollima, A., Pancino, 
E., Ferraro, F.~R., Bellazzini, M., Straniero, O., 
\& Pasquini, L.\ 2005, \apj, 634, 332 

\bibitem[Stanford et al.(2006)]{sdn06} Stanford, L.~M., Da 
Costa, G.~S., Norris, J.~E., \& Cannon, R.~D.\ 2006, \apj, 647, 1075 

\bibitem[Testa et al.(2001)]{testa01} Testa, V., Corsi, C.~E., 
Andreuzzi, G., Iannicola, G., Marconi, G., Piersimoni, A.~M., 
\& Buonanno, R.\ 2001, \aj, 121, 916 

\bibitem[van Loon et al.(2008)]{vanloon08} van Loon, J.~T.,
Boyer, M.~L., \& McDonald, I.\ 2008, \apjl, 680, L49

\bibitem[Worthey et al.(1994)]{Worthey94}
Worthey, G., Faber, S.~M., Gonzalez, J.~J., \& Burstein, D.\ 1994, \apjs,
94, 687

\bibitem[Xin et al.(2007)]{Xin07} Xin, Y., Deng, L., \& Han, Z. W.\ 2007, \apj, 660, 319


\end{thebibliography}
\end{document}